\documentclass[twocolumn]{svjour3}                     % onecolumn (standard format)
\smartqed  % flush right qed marks, e.g. at end of proof
\usepackage{natbib}
\usepackage{graphicx}
\usepackage{amssymb}
\journalname{General Relativity and Gravitation}
\begin{document}

\title{Probing Yukawian gravitational potential by Numerical Simulations. II. Elliptical Galaxies}

%\titlerunning{Probing Yukawian gravitational potential by Numerical Simulations. I. Elliptical Galaxies}
%
\author{C. S. S. Brandao \and J. C. N. de Araujo}
%
%\authorrunning{Brandao \& de Araujo} % if too long for running head
%
\institute{C. S. S. Brandao \and J. C. N. de Araujo \at
    Divis\~ao de Astrof\'\i sica \\
    Instituto Nacional de Pesquisas Espaciais  \\
    Avenida dos Astronautas 1758 \\
    S\~ao Jos\'e dos Campos - 12227-010 SP - Brazil \\
              Tel.: +55-12-39457200\\
              Fax: +55-12-39456811\\
              \email{claudio@das.inpe.br, jcarlos@das.inpe.br}
}

\date{Received: date / Accepted: date}

\maketitle

\begin{abstract}
Since the Newtonian gravitation is largely used
to model with success the structures of the universe,
such as galaxies and clusters of galaxies, for example,
a way to probe and constrain alternative theories, in the
weak field limit, is to apply them to model the structures
of the universe. We then modified the well known \textbf{Gadget-2} code
to probe alternative theories of gravitation through galactic dynamics.
In particular, we modified the \textbf{Gadget-2} code to probe alternatives
theories whose weak field limits have a Yukawa-like gravitational potential.
As a first application of this modified \textbf{Gadget-2} code we simulate the
evolution of elliptical galaxies. These simulations show that galactic dynamics
can be used to constrain the parameters associated with alternative theories
of gravitation.
% 04.50.+h Gravity in more than four dimensions, Kaluza-Klein theory, unified field theories;
% alternative theories of gravity (see also 11.25.Mj Compactification and four-dimensional models)
% 04.50.Kd Modified theories of gravity
% 45.50.-j Dynamics and kinematics of a particle and a system of particles
%% 07.05.Tp  Computer modeling and simulation
\end{abstract}

\section{Introduction}
\label{intro}
Studying Cosmology is a way to probe the physical origin, the present, and destiny of the Universe. The present Cosmological scenario claims that the Universe has the following composition: $\sim 5 \%$ by barionic matter, $\sim 25 \%$ by dark matter, and $\sim 70 \%$  of a component that works like an antigravity term in the Einstein's Equations: the dark energy.

Dark matter can be explained by physically acceptable arguments (e.g., WIMPs or modified theory of gravitation). Forthcoming experiments to be made in the Large Hardron Collider (LHC) will certainly help, in the near future, to give some answers concerning this questions (whether the WIMPs be detected or not).

On the other hand, the dark energy composition is harder to explain, because it can not be thought as due to particles, and its interpretation through a quantum vacuum gives a very bad answer: the energy scale is more than a hundred orders of magnitude greater than the observed value (about this  discrepancy, see, e.g., \cite{cpt1992}). It is very interesting to note that scalar field explanations - such as in some quintessence models - do not solve the problem at all, because other questions arise concerning the very origin of quintessence and its nature.

In this puzzling scenario, some physical theories have been created to explain alternatively the dark energy problem and the dark matter nature. These alternative approaches are not nonsense, because until now, dark matter can only be detected indirectly \textit{via} observations of galactic or extragalactic gravitational interactions, unless LHC changes this situation. These theories are based on different hypothesis (e.g., scalar-tensor theories of gravity,  massive gravitons, etc.) and they do not have in the weak field limit the Newtonian gravitational potential (see, e.g., Refs. \cite{moffat0,moffat1,moffat2,moffat3,moffat4,piazza03,rodriguez-meza05,sign05,araujo2007}).

In some cases, the gravitational potential is Yukawa-like (hereafter Yukawian gravitational potential, YGP). Considering the most simple form of this potential, if we have a point mass $m$, namely

\begin{equation}
\label{ygp}
\phi=-\frac{Gm}{r}e^{-r/\lambda},
\end{equation}
where $r$ is the distance from the point mass, and $\lambda$ is a characteristic length, that means, in some theories, the Compton wavelength of the exchange particle of mass $m_{\rm g}$, a  massive boson called graviton.

It is worth noting that, in almost all works concerning the YGP, the investigations have been made under analytical or numerical approaches. Graviton masses are estimated over many ranges, in different scales of observations, based on different scenarios, from planetary to extragalactic scales. Besides, statical models are commonly used, without taking into account any additional investigations concerning, for example, secular dynamics of the systems.

In this way, we propose numerical simulations of triaxial systems, to probe the YGP, as given by Eq. \ref{ygp}; and investigate how different the galaxies would appear if the YGP were more ``realistic'' than the Newtonian potential at large scales. In other words, if YPG can not keep the King Sphere in dynamical equilibrium, for example, it must be ruled out, because it would destroy the Hernquist spheres and the exponential disks as well, consequently galaxies could not exist in this scenario.

The paper is organized as follows: In Section 2, we present the code and the galactic model used, in Section 3, the simulations and results and, finally, in Section 4, we discuss the results and show the perspectives.

\section{Method and Scenario}
\label{metodo}

As discussed in the previous section, one question to be made is: what does happen to galaxies under the YGP? We know that, from 1970's to now, numerical simulations of galaxies, galactic clusters, and cosmological volumes are used to probe many dynamical and observational features of these astrophysical objects, helping to understand the Cosmos. For example, it is very suggestive that disk galaxies mergers can give rise to giant elliptical galaxies (e.g., \cite{barnes1986}); disk galaxies, their dynamical properties and systems composed by binary disk galactic interactions are probed with numerical simulation tools (see, e.g,  \cite{hernquist1993}, \cite{springel1999} and \cite{springelmatteo2005}), and so on.

In our studies, to simulate an elliptical galaxy under YGP, we have chosen the Gadget-2 code \cite{springel2005} and changed its structure to include the YPG, as we show in detail in our previous work \cite{brandaoearaujo2009a}. With this code, we can integrate all equations of motion of a set of $N$ collisionless particles and follow their evolution.

Gadget-2 is based on the tree code method. So, its computational effort is $N log(N)$, against $N^2$ operations required by direct sum algorithms.

Once a code is chosen, modified and tested, we only need a typical elliptical galaxy model. This is standard matter in galactic dynamics. For our purposes, we choose the King-Michie spheres \cite{bt2008} for two reasons: (I) Although the King spheres have a flat core profile, differently from the cusped core profile given by the Hernquist spheres \cite{hernquist1990}, we are not concerned in reproducing all the observational features from the core of the simulated objects. The overall density profile is sufficient to investigate the YGP. (II) King Spheres are built also in dynamical equilibrium.

King spheres are known as lowered isothermal spheres, idealized to be spherical systems (globular clusters) in equilibrium with the tidal field produced by dark matter halos.  Following  \cite{bt2008}, the distribution function in the phase space of the King sphere is given by
\begin{equation}
f(\varepsilon) = \left\{\begin{array}{rcl} \rho_{1}\frac{1}{\sqrt{(2\pi\sigma^2)^3}} (e^\frac{\varepsilon}{\sigma^2} - 1)
& \mbox{if} & \varepsilon > 0\\
 0 & \mbox{if} & \varepsilon \leq 0, \end{array} \right.
\label{king}
\end{equation}
where
$$f(\varepsilon) \equiv \frac{dN(\varepsilon)}{d^3\vec{x} d^3\vec{v}}$$

\noindent is the number of particles per six dimensional phase space volume around the position $\vec{x}$ and velocity $\vec{v}$; $\rho_1$ is a characteristic density, $\sigma$ is the velocity dispersion of the particles and $\varepsilon = \Phi_0 - \Phi- \frac{1}{2} v^2$, the relative energy, $\Phi$, the system's gravitational potential per unit of mass, $v$, the velocity and $\Phi_0$, a constant, such that $f>0$ for $\varepsilon >0$ and $f=0$ for $\varepsilon \le 0$.

It is important to define the King radius (usually called core radius), namely

$$r_0=\sqrt{\frac{9 \sigma^2}{4 \pi G \rho_0}} \, ,$$

\noindent where the subscript 0 denotes the values of the core.

The King spheres are built with gravitational bound particles, where all particles have velocities $v < v_{esc}$, with $v_{esc}$ being the scape velocity. When $\Phi_0 - \Phi = 0$, $f$ vanishes, and the sphere is truncated, as defined in Eq. \ref{king}. In this case, we define the tidal radius, $r_t$, where the distribution of particles disappears. To King spheres, it is  also useful to define the concentration parameter $\rm{c\equiv log_{10}(r_t/r_0)}$, that estimates how centrally concentrated a model is.

It is worth mentioning that to make the King spheres, one must integrate Eq. \ref{king} to obtain the density distribution at any radius and solve its correspondingly Poisson's equation.  To do this numerically, it is necessary to know the following set of typical parameters: $\rm{W_0}=\frac{\Psi_0}{\sigma^2}$, where $\Psi_0=\Phi(r_t)-\Phi(0)$, $\rm{r_t}$, $\sigma$, $\rm{r_0}$ and $\rm{M_t}$, the total mass of the galaxy.

In the present study we assume the following set of typical parameters for the elliptical galaxies, namely: $\rm{W_0}=7.17$, $r_t$= 79.4 kpc, $\rm{r_0}$= 2.0 kpc, $\sigma$=237.0 $\rm{km.s^{-1}}$, $\rm{M_T}=5 \times 10^{11}$ $\rm{M_{\odot}}$, as can be deduced from observational studies of elliptical galaxies and numerical models of King spheres \cite{bm1998,bt2008}.

\begin{figure}
\center\includegraphics[scale=0.4]{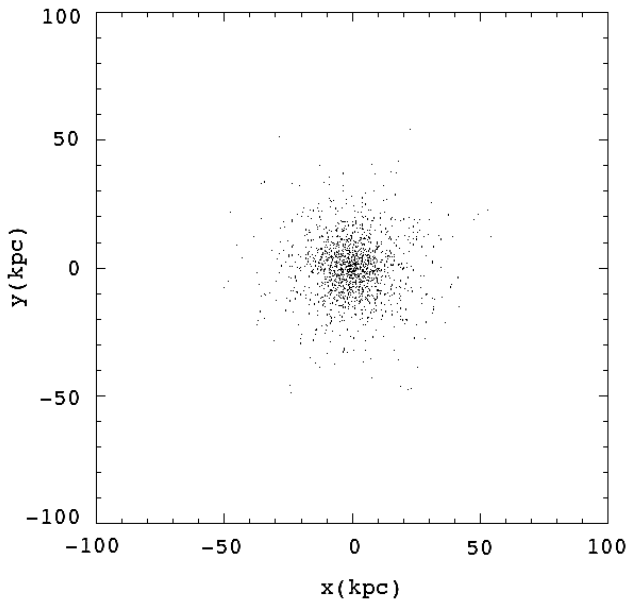}
\caption{Particles's positions of the initial snapshot plotted in the $xy-$plane.}
\label{snap_init}
\end{figure}

To build our galaxy model, we used $N$=2500 particles. We have also made some simulations with more resolution, i.e., using $N$=10000 particles, but the results are very similar, if compared with their respective models in small resolution. So, we decided to maintain small resolutions, turning the analysis procedures easier.  For our purposes, this number is sufficient to map the whole matter distribution and to probe the influence of YGP over the galaxy, including dark and baryonic matters. In Fig. \ref{snap_init}, we show the particles's positions, in the $xy-$plane. Note that the center of the galaxy is located at the origin (0,0,0) kpc.

Our elliptical galaxies are constructed initially with a Newtonian potential and afterwards we submit them to YPG. This procedure could seem misleading, because one could think of we should make galaxies directly with the YGP.  However, it is important to bear in mind that particles represent physical observable quantities, such as positions, velocities and masses distributed over volume. In this philosophy, when we build a initial galaxy snapshot with Newtonian potential, we mimic the following \textbf{observational} characteristics: radial luminosity profile, radial density profile and velocity dispersions. Our particles are statistical sampling from the reliable density distribution, as we expect from collisionless  simulations. In this way, independently from the physics used to build galaxies,  simulated particles must reproduce the observed characteristics from true objects. When the galaxy is then
submitted to the YPG it has its observational characteristics adjusted to this potential. Our aim is then to check,
at the and of our simulations, if these characteristics are consistent with true objects.

\section{Simulations and Results}

\begin{figure}
\center\includegraphics[width=84mm]{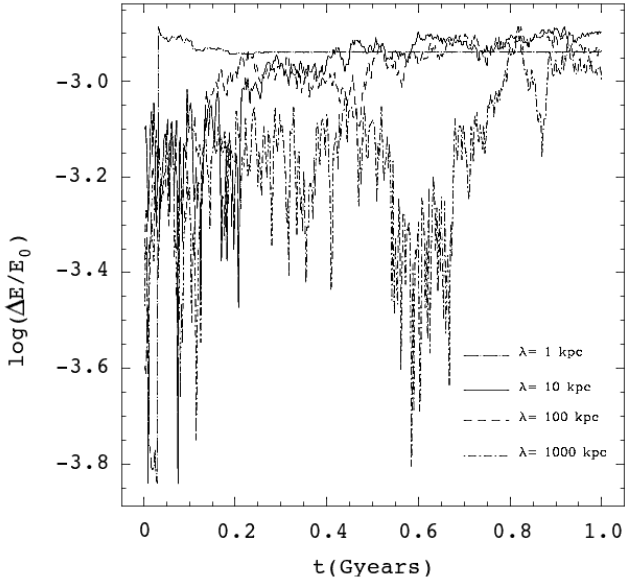}
\caption{Logarithmic values of $\Delta E/E_0$, where $E$ is the total energy at time $t$
and $E_0$, the total initial energy. Time is given in Gyears.}
\label{energy_viol}
\end{figure}

Using our modified \textbf{Gadget-2} code, we performed a set of four numerical simulations to study the behavior of our King spheres under the YPG.  All the simulations have been made with the same galaxy initial conditions, as we explained above, and the same computational parameters, e.g., number of snapshots, initial and final simulated time,  time per snapshot, system of units, etc. We have changed only the parameter $\lambda$, probing its influence in the results.
The principal parameters were fixed as follows: the tolerance parameter $\theta=0.8$, the smoothing length parameter $l$ = 0.1 kpc, to match our maximum simulation resolution with the code capabilities of compute the YPG with small errors, as we have discussed in previous works \cite{brandaoearaujo2009a}. The Yukawa parameter is set as  $\lambda=$1.0, 10.0, 100.0 and  1000.0 kpc, for each simulation, respectively.
Our runs begin at $t=0$ and finish with $t$= 1.0 Gyear. The energy conservation violation is showed if Fig. \ref{energy_viol}. In this picture, we plot the logarithmic values of $\Delta E / E_0$, where $\Delta E = E - E_0$, $E$ is the total energy at time $t$ and $E_0$, the total initial energy. From this figure, we can conclude that the energy violation is smaller than $10^{-2} E_0$, so our simulations are dynamically reliable.

\begin{figure}
\center\includegraphics[width=84mm]{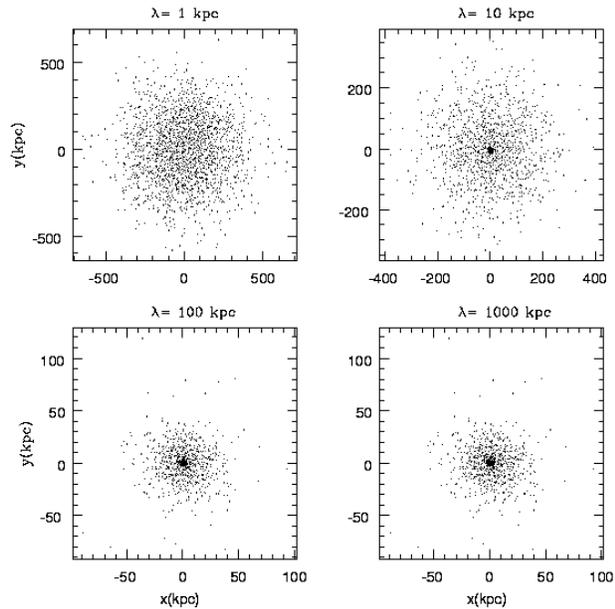}
\caption{ Particles's positions of final snapshots plotted in the $xy-$plane. Boxes are plotted with different scales and each box indicates one type of simulation. The $\lambda$ values are indicated over its correspondingly box.}
\label{snaps_king}
\end{figure}

To see how the galaxies are at the end of the simulations, we display in Fig. \ref{snaps_king} the snapshots after 1 Gyear of simulated time.  This figure shows a bizarre characteristic of this potential for small values of $\lambda$: the galaxy dissolves and sprays over the extragalactic environment. Probing the snaphosts directly, we can observe that the simulated galaxy evaporates faster than it would do in a secular evolution: its core is destroyed at t $\lesssim$ 100 Myears for $\lambda=1$ kpc. In other words, galaxies would not exist. If $\lambda = 10 $ kpc, the galaxy core behaves like a classical object, and can survive for many orbital periods, but regions beyond $r \sim 10$ kpc expands, and the system grows until reach $400$ kpc. As a conclusion, if $\lambda=1$ kpc, we could not see typical galaxies in the cosmos. If $\lambda = 10$ kpc, galaxies would appear fainter, greater and with denser cores, as we will see below.

In cases where $\lambda=100$ kpc,  our triaxial simulated systems can survive for many orbital periods and look like observed systems. In this way, our simulations suggest that, if the YGP were a viable potential, $\lambda \sim 100 $ kpc, otherwise galaxies could not exist, at least typical elliptical systems with $r_t > 10$ kpc or giant cDs.

As we can see in Fig. \ref{snaps_king}, similar aspects of the simulated galaxy are found for $\lambda=10,100$ and 1000 kpc at 1 Gyear. So, to set up the upper limit for the graviton's mass, it is necessary to probe the physical information available in the snapshot files at 1 Gyear and to calculate some observational features of the simulated galaxy. These information are compared to the initial snapshot to see how much the system shifts from its initial values.

\begin{figure}
\center\includegraphics[width=84mm]{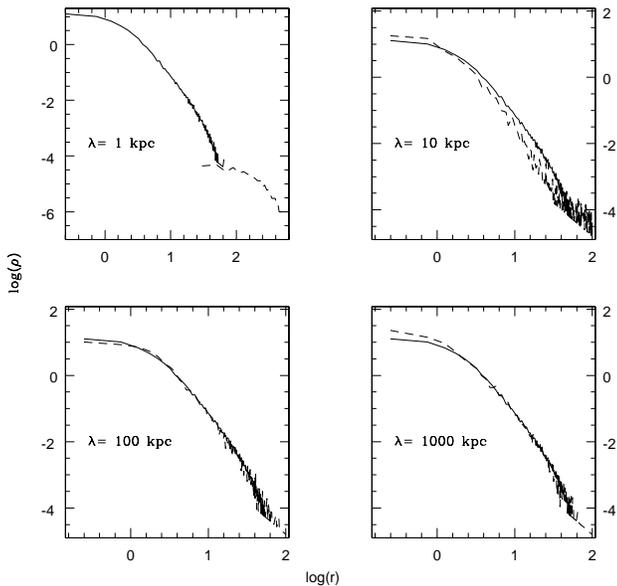}
\caption{Radial density profile of the models. The plot is in log scale. Continuous line represents the density profile for the time $t$ = 0, and the dashed, the final profile, when $t$ = 1.00 Gyear. Each $\lambda$ parameter value is indicated in the boxes.}
\label{density_profs}
\end{figure}

A first and interesting test is to compare the initial radial density profile to the final one. In Fig. \ref{density_profs}, we show the radial density profiles $\rho(r)$ of our simulated  models, where $r$ is the radial coordinate. In this picture, we show solid lines representing the initial data, while dashed lines indicate final snapshots values.

As can be seen in Fig. \ref{density_profs}, the density profile calculated from the simulations for $\lambda$=1.0 kpc lost completely its initial characteristics: due to the fact that particles are lost to extragalactic environment, and a very low density region appears for $ r\lesssim 30 kpc$.

If $\lambda$ = 10.0 kpc, as the time goes on, the sphere grows up, but particles are not lost. The galaxy reassembles its initial profile, but the density profile slope becomes stepper when compared to its initial value.

If $\lambda > 100 $ kpc, the density profile maintains its initial slope. We note that the particle noise effects disturb mostly the right end of these curves and the core densities are affected as well, but these curves conserve the King curves characteristics, despite of the noise effect observed.

Analyzing the Figures \ref{snaps_king} and \ref{density_profs}, it is hard to believe that $\lambda <$  10.0 kpc would be a reliable value. The galaxy evaporates faster than its secular evolution, as a result only the cores could exist. As we can see in the figures, only the cases for $\lambda$=10,100 and 1000 kpc would give reliable results and the models would look like very similar, due to the fact that the density profile resembles its initial value over the entire radial direction.

\begin{figure}
\center\includegraphics[width=84mm]{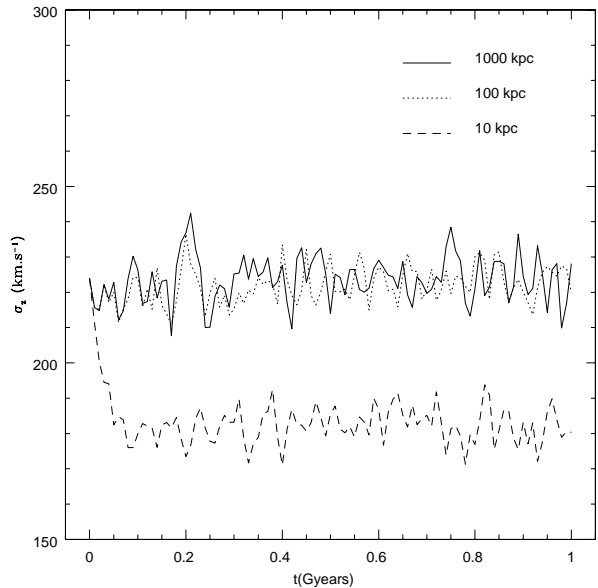}
\caption{Time evolution, in Gyears, of the dispersion velocity in $z-$direction, $\sigma_z$, in km.s$^{-1}$, for different values of the $\lambda$ parameter, namely, 10, 100 and 1000 kpc.}
\label{dispersions}
\end{figure}

Another dynamical test of interest is to measure the dispersion velocity $\sigma$ of the galactic core from the snapshots. The evolution of $\sigma$ can show us how much our models shifts from observed values, i.e., whether the model is reliable or not and how it shifts from the Newtonian behavior. For our purposes, it is sufficient to measure the projected core in a arbitrary axis, resembling a galaxy seen from any projection, like truly observed objects. So, we have measured the $\sigma_z$, the velocity dispersion in the $z-$axis, while the particles are disposed in the $xy-$plane. Any projection would be valid due to the symmetry of King models, so this arbitrary choice is also valid. In Fig. \ref{dispersions}, we show the time evolution of the $\sigma_z$.

Fig. \ref{dispersions} reveals that, when $\lambda=10$ kpc, the core becomes ``colder'' than its initial value. As time goes on $\sigma_z$ oscillates around $\sim 180$ $\rm{km.s^{-1}}$.  This kind of oscillation is expected from numerical  simulations due to particle noise, and happens in other simulations too, as we can be seen in Fig. \ref{dispersions}.

In models with $\lambda \ge 100$ kpc,  $\sigma_z$  oscillates around the initial values, like in Newtonian systems. The Figures and the above analysis show that if the YGP were the true potential, the best Yukawa parameter might be $\lambda > 100$ kpc. From the simulations, it is clear that, as we take small $\lambda$ values, galaxy cores become gravitationally unbound and, in this way, they could not exist, i.e.,  galactic cores became colder until disappear, as we saw in Fig. \ref{snaps_king}.  Under this physical requirement, we can fix an upper limit for the graviton mass: $\rm{m_g} \ll 10^{-60}$ g.

%\section{Conclusions}
\section{Discussions and conclusions}

In this work, we have studied triaxial systems to probe the YGP and to constrain the yukawian $\lambda$ parameter. We showed here that, if YGP were reliable, we should have $\lambda > 100$ kpc. This value is larger then that from solar system's constraints and it must be considered a good estimative, since with such a value the simulated galaxies remains ``alive" for billions of years and look like their observational counterparts. It is worth stressing that triaxial systems easily constrain the $\lambda$ parameter, because we only need to study the radial density profile, particles's positions and the core's velocity dispersion. Although this procedure is somewhat simplified, it give us sufficient information about the status of the model's structure and its observational counterparts. In particular, models presenting the destruction of a galaxy, such that depicted in the first frame of Fig. \ref{snaps_king}, or presenting bizarre morphologies mean that the parameter under consideration is unreliable.

This work has some important characteristics. We have designed a method to probe alternative theories of gravitation in the \textit{non relativistic r\'egime}, simulating ``alive" galaxies submitted to the investigated potential. This method reveals itself as a test to constraint the parameters of the probed theory. Concerning the alternative theories of gravitation, we have developed a pioneer simulation method probe in the sense that previous works have studied statical models (e.g., using galactic scenario, de Araujo \& Miranda studied disk galaxies under YGP, but using analytical arguments), while our galaxies behave like ``alive" systems, because are composed by thousands of gravitating particles. In this way, this must be considered a reliable and strong test, due to the fact that N-Body systems are very sensitive to the physics used in the simulation, and their observational counterparts are required to give us the best parameters of the potential under investigation. Although chaos and complex phenomena appears in the N-Body systems, it is important to bear in mind that these systems are well understood by Galactic Dynamics (see, e.g., \cite{bt2008}).

Another interesting test would be making disk galaxies under the YGP. This test would be a more sophisticated one, due to the fact that late-type systems have complex substructures that are expected to appear in these simulations, like spiral arms, bars and rings. These features can be used to probe the $\lambda$ values with more precision. In other papers to appear elsewhere \cite{brandaoearaujo2009b}, \cite{brandaoearaujo2009c} we will make these simulations under YGP and other potentials.
\begin{acknowledgements}
CSSB and JCNA would like to thank the Brazilian agencies CAPES and FAPESP for support. JCNA would like also to thank the Brazilian agency CNPq for partial support.
\end{acknowledgements}

\end{document}